\documentclass[
          twoside=false,
          twocolumn=true,
          a4paper,
          10pt]{scrartcl}

\usepackage[top=1.5cm, bottom=2.5cm, left=1.5cm, right=1.5cm]{geometry}

\usepackage{ifpdf}

\ifpdf
   \usepackage[pdftex]{graphicx}
   \DeclareGraphicsExtensions{.pdf,.jpeg,.png}
\else
  \usepackage[dvips]{graphicx}
  \DeclareGraphicsExtensions{.eps}
\fi

\usepackage[caption=false,font=normalsize,labelfont=sf,textfont=sf]{subfig}

\usepackage{flushend}

\begin{document}

\title{Evaluating Connection Resilience for Self-Organized Distributed Cyber-Physical Systems}

\author{
  Henner Heck, Olga Kieselmann and Arno Wacker \\
 \texttt{\{henner.heck$|$olga.kieselmann$|$arno.wacker\}@uni-kassel.de}
}

\date{}

\maketitle

\begin{abstract}

Self-organizing cyber-physical systems are expected to become increasingly important in the context of Industry 4.0 automation as well as in everyday scenarios.
Resilient communication is crucial for such systems.
In general, this can be achieved with redundant communication paths.
Mathematically, the amount of redundant paths is expressed with the network connectivity.
A high network connectivity is required for collaboration and system-wide self-adaptation even when nodes fail or get compromised by an attacker.
In this paper, we analyze the network connectivity of a communication network for large distributed cyber-physical systems.
For this, we simulate the communication structure of a CPS with different network parameters to determine its resilience.
With our results, we also deduce the required network connectivity for a given number of failing or compromised nodes.

\end{abstract}

\section{Introduction}
\label{conn:sec:introduction}

Based on definitions from literature (\cite{sander2013, Lee06, broy2012cyber}), we define a \emph{Cyber-Physical System (CPS)} as a distributed system which integrates computational and physical processes and consists of multiple interconnected nodes.
The main property of such systems is to adapt to changing physical processes, which cannot be predicted entirely at design time.
This makes CPS interesting for the Industry 4.0 context as well as everyday scenarios. In Industry 4.0, highly automated systems need to deal with complex processes taking place in the physical world in an autonomous manner. In \cite{lee2015cyber}, the authors showed that CPS can improve automated processes significantly by exploiting self-* properties.

An example for a CPS is a Smart Camera Network (SCN). Smart cameras are video cameras with a built-in computation unit that can be utilized for various tasks, e.g., image processing, object localization or object tracking. With the integrated wired or wireless communication devices, smart cameras are able to communicate with each other. Today, the most common scenario for a SCN are the detection of intruders in restricted areas or the surveillance of high risk areas. In general, the observed information from single cameras is used to classify processes in the whole environment for assessing its current situation and adapting the behavior of single nodes, e.g., to follow suspects. A detailed smart camera scenario in an organic computing context is presented in \cite{esterle_et_al_2013_camsim}.

Another example for a CPS is a distributed network intrusion detection system (IDS) for securing cooperate networks with several branches. At each branch, multiple nodes observe and classify all passing traffic as \emph{suspicious} or \emph{normal}. This can be based on rules (e.g.,~\cite{snort99}) or a Gaussian mixture model (e.g.,~\cite{Fis12a}). The nodes exchange their observed information regularly to increase the chance of detecting distributed attacks. Such attacks might not be detected by individual nodes, but with the aggregated information the detection possibility increases and allows for faster and more efficient adaptation by each node.

In both given examples, the system must be able to adapt to complex situations in the physical environment going beyond the scope of a single node.
A necessary feature for this is the information exchange about the physical environment between the nodes. This information must be exchanged via communication channels, which can be either direct or indirect via other nodes.
However, we must consider that nodes or communication channels might fail.
Additionally, there exist a number of security weaknesses as described in, e.g.,~\cite{broy2012cyber,Cardenas2008,eckert2013, krueger2013, sander2013}. More specifically, since some nodes might be publicly accessible in the physical world, we must consider that they can be compromised by an attacker.
If a node is compromised, all communication from or through that node must also be considered compromised.
To still achieve reliable self-adaption, we must never rely on information from single nodes, and we require redundant communication channels for resilient inter-node communication \cite{heck2016multi}. More precisely, to tolerate compromised nodes, there must be multiple node-disjoint communication paths through the network for any node pair. The minimum number of node-disjoint paths for any node pair in a network is called the \emph{network connectivity} of the communication network.

Highly distributed self-organizing CPS exhibit coordination schemata and communication requirements which are similar to structured overlay networks.
Therefore, when designing a CPS, we can benefit from proven communication concepts from such networks.
Especially, distributed hash tables (DHT) are suitable for information exchange and storage in CPS.
They avoid centralistic structures and, therefore, bottle necks and single points of failure.
We, therefore, assume that the nodes of the CPS are organized as a structured overlay network, specifically as a DHT.
For our analysis, we chose a DHT formed by Kademlia~\cite{maymounkov2002kademlia} as it is one of the most used protocols. 
Next to receiving much attention in the scientific community, it has been successfully deployed in several real life applications with thousands to tens of thousands of nodes, e.g., BitTorrent~\cite{loewenstern2008bep}.

However, the applicability of Kademlia to a large scale CPS as well as its network connectivity have not yet been researched. A high network connectivity is a requirement for designing a resilient CPS. Therefore, in this paper, we evaluate the network connectivity of Kademlia to determine its general suitability for designing a resilient and self-adapting CPS. We present the results of extensive simulations with different network configurations and measure their effect on the network connectivity (resilience) and recovery (self-healing). 

The rest of this paper is organized as follows: First, we discuss related research about DHT network connectivity in Section~\ref{conn:sec:relatedwork} and present our assumptions in Section~\ref{conn:sec:systemmodel}. After that, we briefly describe in Section~\ref{conn:sec:connectivity} the Kademlia protocol and the mathematical foundations for computing the network connectivity. Based on this, we preset  and discuss the results of our connectivity measurements in Section~\ref{conn:sec:evaluation}. Finally, we conclude our paper in Section~\ref{conn:sec:conclusion} with a brief summary and provide an outlook on future research.

\section{Related Work}
\label{conn:sec:relatedwork}

Kademlia and DHTs in general have been studied extensively in the scientific literature. A survey about research on robust peer-to-peer networks from 2006 \cite{risson2006survey} already lists several hundred references. Another survey from 2011 with focus on security aspects in DHT, reaches close to a hundred references \cite{urdaneta2011survey}.
Despite the large amount of publications in general, the global network connectivity of Kademlia has not been thoroughly evaluated. We limit our discussion of related work to literature with relevance for connectivity of structured overlay networks built with Kademlia or it's descendants.

In \cite{kovacevic2008towards}, the authors simulate Kademlia networks and apply churn (joining/leaving of nodes) to evaluate resilience. While the basic premise is similar to ours, they measure response times and number of message hops, not network connectivity.
In \cite{jimenez2009connectivity}, the authors insert nodes into a real-world BitTorrent network.
From these nodes, they try to make contact with other nodes in the network to assess the general reachability of nodes in that network.
The main focus of this paper is on connectivity problems within the network caused by technical obstacles such as firewalls and network address translation (NAT).
Furthermore, the authors analyze connectivity properties of small groups of nodes such as ``transitivity'' and ``reciprocality''. They do not measure the network-wide connectivity.
Similarly, the authors of \cite{crosby2007analysis} insert nodes into real-world overlay networks built by the BitTorrent protocol to measure round trip times and message rates for resource lookups. Additionally, they measure ``connectivity artifacts'' and ``communication locality''. Artifacts emerge from nodes making contact with the author's nodes, but cannot be contacted by them. As in \cite{jimenez2009connectivity}, the authors conclude that such a behaviour is most likely caused by firewalls and NAT.
The communication locality measurements show to what degree nodes in the DHT preferably communicate with other nodes that, according to the protocol's definition of node distance, are near to them.
While both properties are related to the network connectivity, it is not measured or derived.
The authors of \cite{salah2013capturing} present a crawling software for capturing connectivity graphs of networks built by the KAD protocol, a descendant of Kademlia. They insert specially modified crawling nodes into a real-world network to contact other nodes and dump the contents of their routing tables. The dumped routing tables are then used to create a connectivity graph of the network. Though they capture connectivity graphs, their focus is on the crawling process, and they do not perform any analysis on the captured graphs.
In \cite{baumgart2007s}, the authors propose different measures to make Kademlia networks more resilient towards malicious nodes. One of those measures is the use of node-disjoint paths for lookup procedures. The authors measure success rates for lookup procedures using different numbers of disjoint paths. Their simulations imply that a certain average level of connectivity is present in a network, but they do not measure the actual connectivity.

In contrast, our main goal is to determine the network connectivity of Kademlia in dependence of its parameters. Some of the related work, e.g., \cite{baumgart2007s}, even rely on a given network connectivity, but it was neither determined analytically nor experimentally before.

\section{System Model}
\label{conn:sec:systemmodel}

The CPS is organized as a distributed system consisting of multiple networked \emph{nodes} which interact via \emph{communication channels}. The basic functioning of one node is not dependent on the functioning of others. Each node has sensors and actuators to observe a part of the \emph{physical environment} and to interact with it.

The self-organization and specifically the self-adaptation at runtime is based on the Observer/Controller architecture (e.g., \cite{tomforde2011observation, hahner2013concept}).
To achieve sufficient observation and adaptation and to choose appropriate means of control, multiple nodes must cooperate.
This requires the exchange of information among the nodes.
Therefore, while each node has its own local task, it also communicates and collaborates with other nodes towards a global system goal.
Each node is able to communicate with any other node, either directly or indirectly via others.
The communication structure of the CPS is organized as a structured overlay network, i.e., as a DHT based on Kademlia (cf.\ Section \ref{conn:subsec:kademlia}).

We assume the presence of an \emph{attacker} with the goal of disturbing, disabling or controlling nodes of the CPS.
We call a node which has been successfully attacked a \emph{compromised node}.
There are several other causes exhibiting the same effect as a compromised node.
Without additional measures, these are indistinguishable from an attack, e.g., maintenance, failures from defects, or other disturbances like power outages.
If an attacker has compromised a node, we assume that she is able to fully impersonate the node towards the rest of the CPS.
Therefore, an attacker can disseminate information into the network as a legitimate part of the system and also deny requests coming from other nodes and, thus, hinder or prevent information exchange.
Communication between two nodes is not always direct, so other nodes can be necessary for message transfer.
Therefore, a compromised node includes the case of compromised communication channels.
Additionally, we assume that the attacker can subvert at most $a$ arbitrary nodes at any time.

\section{Connectivity}
\label{conn:sec:connectivity}

In this section, first, we present the properties and mechanisms of Kademlia important for routing and contact management.
To analyze the network connectivity, we introduce the mathematical foundations to transfer the network structure of Kademlia into the domain of graph theory by creating a connectivity graph.
Next, we describe the mathematical algorithms and necessary graph transformations for calculating the graph connectivity. Finally, we use the mathematical foundations to define the resilience of the communication network.

\subsection{Kademlia}
\label{conn:subsec:kademlia}

With Kademlia, each node and each stored data object is identified by a numerical \emph{id} with the fixed bit-length $b$. These identifiers are generated from a node's network address or the data object respectively, using a cryptographically secure hash function with the goal of equal distribution of identifiers in the identifier space.
Each node maintains a routing table with identifiers and network addresses of other nodes, its so-called \emph{contacts}.
The routing table consists of $b$ so-called $k$-buckets to store the contacts of the node.
The buckets are indexed from $0$ to $b-1$, and the contacts are distributed into these buckets depending on the distance of their identifiers $\mathit{id}_i$ and the node's id.
For this, the distance between two identifiers is computed using the XOR metric, meaning that for two identifiers $\mathit{id}_a$ and $\mathit{id}_b$ the distance is $\mathit{dist}(\mathit{id}_a,\mathit{id}_b)=\mathit{id}_a \oplus \mathit{id}_b$, interpreted as an integer value.
The buckets are populated with those contacts $\mathit{id}_i$ fulfilling the condition $2^i \leq  \mathit{dist}(\mathit{id},\mathit{id}_i) < 2^{i+1}$, with $i$ being the bucket index.
This means that the bucket with the highest index covers half of the id space, the next lower bucket a quarter of the id space, and so on.
The maximum number of contacts stored in one bucket is $k$.
Next to $b$ and $k$, the third defining property of a Kademlia setup is $\alpha$, which determines how many contacts are queried in parallel when a node wants to either locate another node or retrieve/store a data object. The Kademlia authors propose to set $b=160$, $k=20$ and $\alpha=3$.

The nodes of a Kademlia network can locate resources (other nodes, data objects) by means of their identifiers. Given a target identifier, a node queries $\alpha$ nodes from its routing table nearest to that targets identifier. In turn, each queried node answers with its own list of nodes closest to the target identifier. This way, the requesting node iteratively gets closer to the target identifier.
This process ends when a number of $k$ nodes have been successfully contacted, or no more progress is made in getting nearer to the target identifier.

In addition to acting as a building block for a larger self-organizing system, we find it remarkable that Kademlia itself also exhibits \emph{self-*} properties. Depending on its own identifier and that of other nodes, each node will build a different routing table, i.e., perform \emph{self-configuration}.
Emerging properties for the overlay network are its connectivity and a small number of hops necessary for locating nodes or data objects.
The protocol enables nodes to detect stale contacts in their routing table and replace them to restore connectivity, thereby exhibiting \emph{self-healing} behaviour.
On replacing a stale node from the routing table, the most recently seen node from a list of possible replacements is chosen. This is done to optimize the chance of a non-stale replacement node. By doing so, the network continuously \emph{self-optimizes} its routing.

\subsection{Connectivity Graph}
\label{conn:subsec:connectivity_graph}

The representation of the network structure as a connectivity graph enables the application of concepts and algorithms from graph theory to analyze properties of the network.
The connectivity graph $D(V,E)$, with the vertices $V$ and edges $E$, is a directed graph representation of the nodes and their routing tables.
Each vertex from the connectivity graph represents a distinct node from the network. Hence, the number of vertices equals the number of network nodes.
To construct the connectivity graph, we add edges the graph according to the routing table of Kademlia.
For each node pair $\mathit{id}_i$, $\mathit{id}_j$ represented in the graph by vertices $v$ and $w$ respectively, we insert the directed edge $(v,w)$ into the set of edges $E$ if and only if node $\mathit{id}_j$ is present in the routing table of $\mathit{id}_i$.

Generally, in network graphs, a capacity value is often assigned to the edges for expressing the communication bandwidth between nodes. This is not a necessity for connectivity graphs, since the existence of the edges is enough to indicate a connection between nodes. However, since it is necessary for later steps, we assign a capacity of $1$ to each edge.

\subsection{Vertex Connectivity for Vertex Pairs}
\label{conn:subsec:vertex_connectivity_pair}

A directed edge in the connectivity graph $D(V,E)$ can be interpreted as an one-way water pipe.
The maximum amount of water able to flow through the pipe per time unit is modeled by the edge capacity.
The \emph{maximum flow} between two vertices $v$ and $w$ is the sum of the capacities of the \emph{minimum edge cut}.
This is the set of edges with the smallest total capacity whose removal would cut off any flow from $v$ to $w$.
In other words, the minimum edge cut is the bottle neck which determines the maximum possible flow $v$ to $w$.

Analog to the minimum edge cut for two vertices $v$ and $w$, the \emph{minimum vertex cut} is the minimum number of vertices whose removal cuts all paths from $v$ to $w$.
The order of the minimum vertex cut is called the \emph{vertex connectivity} from $v$ to $w$, i.e., $\kappa(v,w)$.
Menger's theorem \cite{menger1927allgemeinen} for directed graphs \cite{beineke2013topics} states that for the two non-adjacent vertices $v$ and $w$ the vertex connectivity is equal to the maximum number of pairwise vertex-disjoint paths from $v$ to $w$.
This number correlates directly with the communication resilience (cf.~Section \ref{conn:subsec:resilience}).
Therefore, to evaluate the resilience, we need to calculate the vertex connectivity.

There are multiple algorithms to compute the maximum flow/minimum edge cut between any two vertices in a graph.
However, in general, the vertex connectivity does not correspond to the maximum flow/minimum edge cut.
To bridge the gap from computing the maximum flow/minimum edge cut to computing the vertex connectivity, we apply Even's algorithm (e.g., \cite{even1973algorithmic, even1975algorithm, even1979graph, even2011graph}).
It transforms the connectivity graph $D(V,E)$ such that the maximum flow between two non-adjacent vertices is equal to their vertex connectivity.
This allows the application of maximum flow algorithms to calculate the vertex connectivity.
Even's graph transformation is applied on the original connectivity graph $D(V,E)$ consisting of $n$ vertices and $m$ edges. We assume that $D(V,E)$ has neither self-loops nor parallel edges.
The problem transformation is done by applying the following steps to each vertex of $D(V,E)$:

\begin{itemize}
\item Let $s$ be a vertex of the directed graph $D(V,E)$ with the incoming degree of $d_{\mathit{in},s}$ and outgoing degree of $d_{\mathit{out},s}$.
\item Split $s$ into the two vertices $s'$ (incoming vertex) and $s''$ (outgoing vertex).
\item All incoming edges of $s$ point to $s'$, so that it has the incoming degree $d_{\mathit{in},s}$.
\item Make all outgoing edges of $s$ originate from $s''$, so that it has the outgoing degree $d_{\mathit{out},s}$.
\item Insert the edge $(s',s'')$ with capacity $1$, so that the outgoing degree of $s'$ and the incoming degree of $s''$ are both $1$.
\end{itemize}

\begin{figure}
\subfloat[Original Network Graph $D$.]{\includegraphics[width=\columnwidth]{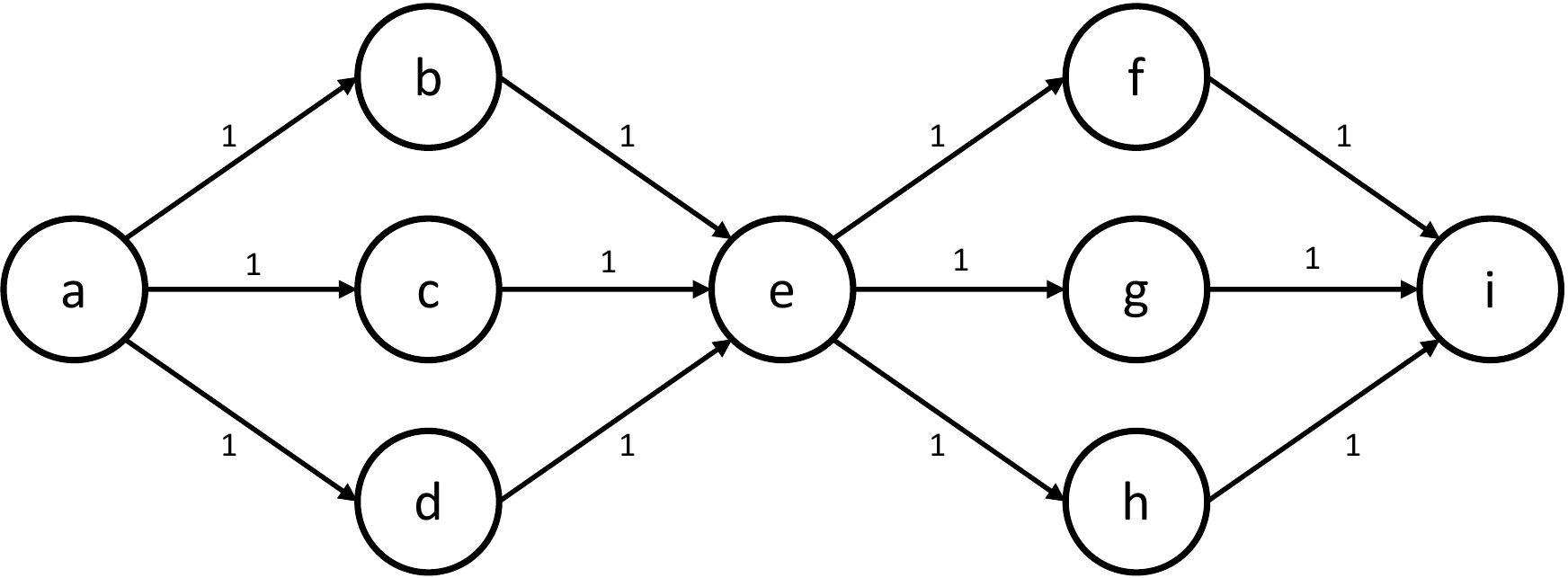}} \qquad \qquad \qquad
\subfloat[Transformed Network Graph $D'$.]{\includegraphics[width=\columnwidth]{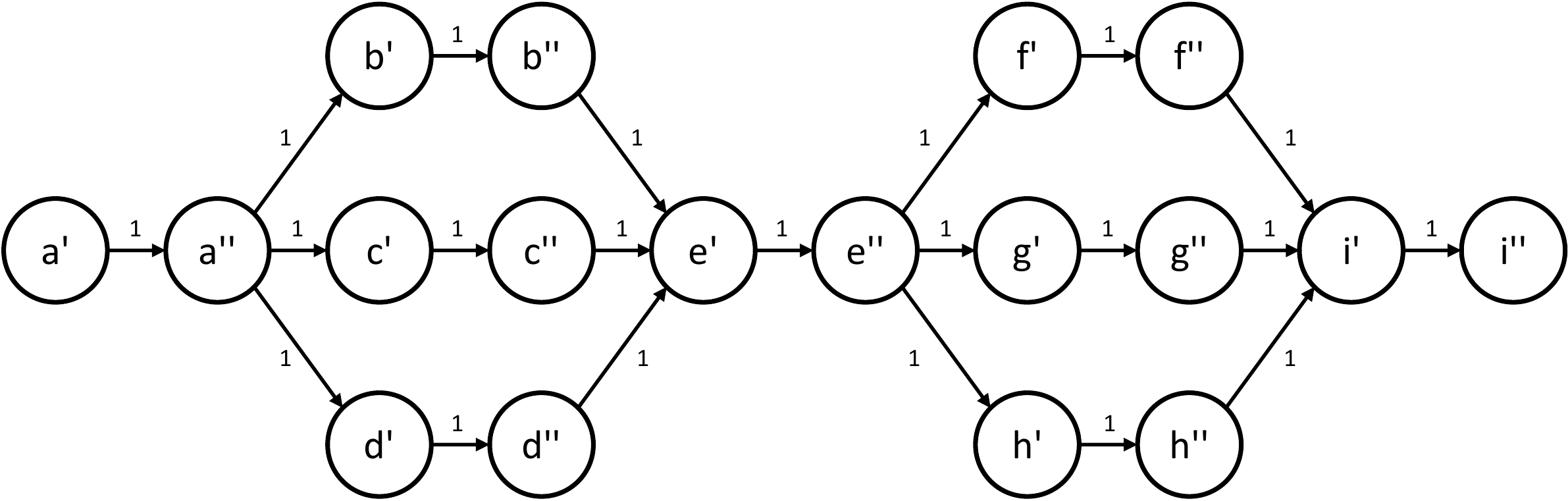}}
\caption{Example graph transformation for Even's algorithm. From vertex $a$ to vertex $i$, the connectivity graph in (a) shows a maximum flow of $3$ and a vertex connectivity $\kappa(a,i) = 1$. For $a''$ and $i'$ in the transformed graph $D'$ in (b), the maximum flow equals the vertex connectivity of $1$.}
\label{even_graphs}
\end{figure}

The resulting graph $D'(V',E')$ has $2n$ vertices and $m+n$ edges and can be used to calculate the vertex connectivity by applying a max flow algorithm.
An example for such a graph transformation is shown in Figure \ref{even_graphs}.

\subsection{Vertex Connectivity for Graphs}
\label{conn:subsec:vertex_connectivity_graph}

The vertex connectivity of a graph $D(V,E)$ is the minimum of the vertex connectivities of all pairs of distinct non-adjacent vertices in the graph, i.e.,
\begin{equation}
\kappa(D)=\mathit{min}(\kappa(v,w)) : v \neq w \wedge (v,w) \notin E \wedge v,w \in V.
\end{equation}
If $D(V,E)$ is not a complete graph, we determine the vertex connectivity $\kappa(v,w)$ for a pair of non-adjacent vertices $v$ and $w$ by computing the maximum flow from outgoing vertex $v''$ to the incoming vertex $w'$ in the transformed graph $D'(V',E')$. Therefore, the vertex connectivity $\kappa(D)$ for the whole graph can be determined by finding the minimum of the maximum flows between all pairs of outgoing and incoming vertices in the transformed graph $D'(V',E')$.
If $D(V,E)$ is complete, meaning that any vertex is adjacent to any other vertex, the vertex connectivity is the number of vertices in the graph $n$ minus one \cite{beineke2013topics_ch0}.

To find the minimum of the maximum flows for a transformed directed graph $D'(V',E')$ with $2n$ vertices, it is generally necessary to compute the maximum flow for all $n(n-1)$ distinct pairs of outgoing/incoming vertices. This makes the time complexity in terms of maximum flow computations $\mathcal{O}(n^2)$.

To find the minimum of the maximum flows for a transformed undirected graph $G'(V',E')$ with $2n$ vertices, it is sufficient to compute the maximum flow for $n-1$ distinct pairs of outgoing/incoming vertices \cite{gomory1961multi}. This makes the time complexity in terms of maximum flow computations $\mathcal{O}(n)$.

\subsection{Resilience}
\label{conn:subsec:resilience}

As stated in our system model (Section \ref{conn:sec:systemmodel}), we assume that an attacker is able to subvert a number of $a$ nodes of the CPS.
We require that information exchange in the CPS is still reliable even under this condition.

We call a CPS that can function properly even when a number of $r$ nodes have been subverted, an $r$-resilient CPS.
This means that with $r$ subverted nodes a reliable path must still exist between any pair of nodes in the network.
Hence, to tolerate $a$ compromised nodes, we need an $r$-resilient CPS with $r \geq a$.
This is fulfilled when the connectivity $\kappa(D)$ is greater than the number of compromised nodes, i.e., $\kappa(D) > r$.
Since each compromised node can disconnect at most one of the $\kappa(D)$ node-disjoint paths (cf.\ Section \ref{conn:sec:connectivity}), there is still at least one reliable path remaining.

Therefore, the correlation between the graph connectivity, the resilience and the number of attackers can be summarized in Equation \ref{con:eq:resilience}.
\begin{equation}\label{con:eq:resilience}
  \kappa(D) > r \geq a
\end{equation}
From this equation, we are able to determine (1) the resilience of a CPS for a given network as $r=\kappa(D)-1$ and (2) the required connectivity of a network for a certain $a$ as $\kappa(D) > a$.

\section{Evaluation}
\label{conn:sec:evaluation}

In this section, we first describe our simulation environment, i.e., the tools used to determine the network connectivity. After that, we present our evaluation methodology and the simulation scenarios. Finally, we present the achieved results and discuss them.

\subsection{Environment}
\label{conn:subsec:environment}

For our simulations, we use the network simulation software PeerSim \cite{montresor2009peersim}.
It is implemented with the Java programming language and includes an event protocol class for event driven simulations. We added Kademlia as an instance of this ``EDProtocol''.
Additionally we wrote software components to provide functionality for creating network churn (addition and removal of nodes) as well as requesting data objects and disseminating information into the network.

For the graph transformation, we implemented Even's algorithm in Java.
To calculate the maximum flow between a pair of vertices, we use the software ``HIPR'' \cite{goldberg_hipr}. It is a C implementation of the hi-level variant of the push-relabel algorithm presented in \cite{cherkassky1995implementing}.
In its original form, HIPR only calculates the maximum flow for one vertex pair.
Therefore, we modified it to support calculations with multiple vertex pairs per program invocation.
As adjacent vertex pairs do not influence the graph connectivity in our context (cf.~\ref{conn:subsec:vertex_connectivity_pair}), we also added program logic to detect and skip such pairs.
We further wrote multiple software tools and scripts for both generation of maximum flow computing tasks, and validation and aggregation of the output created from these tasks.

We ran our simulations on two computers each with an Intel i7 quad core CPU with hyper-threading. For the maximum flow computations, we used a Linux cluster provided by our University. We distributed the computations to 24 cluster nodes each providing two 16 core AMD Opteron 6276 CPUs (2.3 GHz) with hyper-threading.

\subsection{Methodology}
\label{conn:subsec:methodology}

To calculate the graph connectivity over time, we persist the connectivity graph of a network at pre-defined time stamps in a simulation.
For that purpose, we interrupt the simulation and save the current contents of the routing tables of all network nodes to disk into a snapshot file.
We use this snapshot file to transform the connectivity graph with Even's algorithm.
Next, we convert the transformed graph to the supported input format of HIPR (i.e., DIMACS \cite{dimacs}) to calculate the maximum flow.

The push-relabel algorithm used for the maximum flow computation for a single vertex pair in HIPR has a worst case time complexity of $\mathcal{O}(n^2 \sqrt{m})$, where $n$ is the number of vertices and $m$ the number of edges in the processed graph \cite{cherkassky1995implementing}.
Since the transformed graph $D'(V',E')$ contains $2n$ nodes and $n+m$ edges, the complexity of calculating the maximum flow of a single vertex pair in $D'$ is $\mathcal{O}(n^2 \sqrt{n+m})$.
To calculate the graph connectivity $\kappa(D')$, we need to apply the above calculation on the transformed graph from all outgoing vertices to all incoming vertices, i.e., $n(n-1)$ times.
Thus, the overall time complexity for calculating $\kappa(D')$ is $\mathcal{O}(n^4 \sqrt{n+m})$. This complexity makes the maximum flow computation very expensive. For instance, the full maximum flow computation for a transformed connectivity graph with $2500$ vertices takes about 125 hours on a single core of our hardware.

The nodes in Kademlia attempt to add each other to their respective routing tables. This would result in an undirected connectivity graph.
However, due to size restrictions of the buckets in the routing table and race conditions, these attempts are not always successful.
Hence, there is no guarantee for the connectivity graph being undirected.
Nevertheless, our analysis of simulation runs shows that the connectivity graphs come very close to being undirected.
This allows us to reduce the amount of maximum flow computations from $n(n-1)$ to $c \cdot n(n-1)$, $0 < c \leq 1$.
We achieve this reduction by only using a percentage $c \cdot n$ of outgoing vertices for the maximum flow calculation.
Since the outgoing degree $d_{\mathit{out},v}$ of a vertex $v$ is an upper limit for the outgoing flow, we select those $c \cdot n$ outgoing vertices with the smallest $d_{\mathit{out}}$.
As we calculate the maximum flow from only a percentage $c \cdot n$ of outgoing vertices to all $n-1$ incoming vertices, also the limiting incoming degree $d_{\mathit{in}}$ is still considered.
We verified this with 20 randomly selected connectivity graphs, for which we performed a full analysis, i.e., calculated the maximum flow for all $n(n-1)$ vertex pairs.
In all 20 cases, $c=0.02$ (2\%) was sufficient to determine the minimum of the maximum flows, i.e., the graphs vertex connectivity.

\subsection{Scenarios}
\label{conn:subsec:scenarios}

To determine which environment and Kademlia parameters influence the connectivity of the network we devised five dimensions for the simulations, i.e., network size, network setup, network churn, network traffic,  and the Kademlia bucket size $k$.

\subsubsection*{Network Size}

We consider two different scenarios for the network size, i.e., a network with 250 nodes and one with 2500 nodes.
Our choice for these network sizes is closely related to real world CPS' given in Section \ref{conn:sec:introduction}.
For the smart camera scenario, a large number of smart cameras may be necessary for reliably observing and controlling a large industrial complex. Thus, we simulate it with $250$ nodes.
In contrast, a distributed IDS can be used for securing corporate networks spanning over several branches.
Such networks usually comprise several hundreds to thousands of nodes. Exemplarily, we choose $2500$ nodes for this scenario.

\subsubsection*{Network Setup}

The initial bootstrap procedure to create the network is done sequentially. A new node joins every 180 milliseconds until the intended network size is reached.
We distinguish three different scenarios for setting up the network, i.e., which nodes are used as bootstrap nodes.
In the first network setup, a new node chooses the bootstrap node randomly from all already joined nodes -- we call this the random scenario (\emph{R}).
In this scenario, the simulation may select any node to be removed from the network during simulation runs with churn.
The second network setup is the same as before, but there is a group of five nodes in the network which will never be removed during simulation runs with churn. We call this the stable scenario (\emph{S}).
Finally, in the third network setup, a new node always selects one of the stable nodes as its bootstrapping node. i.e., the bootstrapping scenario (\emph{B}).
We name the average connectivity of all measured node pairs \emph{Ravg}, \emph{Savg} or \emph{Bavg}. We name the minimum connectivity of the network \emph{Rmin}, \emph{Smin} or \emph{Bmin}.

\subsubsection*{Network Churn}

We consider five different churn scenarios.
In the first scenario, we do not consider any churn, i.e., no new nodes join, and no nodes are removed.
In the second scenario (0/1), we remove a single node from network in every minute and add no nodes.
Similarly, we add one node and also remove one in every minute in our third scenario (1/1).
The last two scenarios are so-called burst scenarios.
In the forth scenario (0/19), we remove 19 nodes simultaneously (burst) every 10 minutes.
We chose a burst of 19, as it is slightly less than the default bucket size for Kademlia ($k=20$).
Finally, in the fifth scenario (0/40), we simultaneously remove 40 nodes every 10 minutes.
We chose 40 to determine the impact of concurrently removing significantly more nodes than the bucket size.

\subsubsection*{Network Traffic}

We distinguish two different scenarios with respect to data traffic, i.e., with and without data traffic.
In the scenario with data traffic, all nodes regularly look up data objects and disseminate them.
For this, each node performs $10$ lookup procedures an $1$ dissemination procedure per minute during the whole simulation.
In the scenario without data traffic, the node do not lookup data objects or disseminate them.
However, for maintenance purposes Kademlia requires each node to perform a so-called ``bucket-refresh'' every 60 minutes.
For this, each node randomly selects ids from its contacts and performs lookup procedures for these ids.
This way, it can learn about previously unknown contacts and stale contacts in its routing table.
Hence, even in the scenario without data traffic, there is some basic background traffic.

\subsubsection*{Kademlia Bucket Size}

In Kademlia, the bucket size $k$ is directly responsible for the number of contacts a node can keep in its routing table.
To determine its effect on the network connectivity, we differentiate four different values for $k$, i.e., $k\in\{5,10,20,30\}$.

In summary, we have five dimensions with several scenarios for each of them, i.e., $2\cdot 3 \cdot 5 \cdot 2 \cdot 4 = 240$ combinations. We simulated all combinations to determine how the dimension and the connectivity correlate. We present the results of selected simulations in the next section.

\subsection{Results}


In this section, we first present the simulation and measurement results for a network size of 250 nodes (Simulation 1-5) and 2500 afterwards (Simulation 6-11).

\subsubsection*{Simulation 1}

\begin{figure*}
  \centering
  \includegraphics[width=\textwidth]{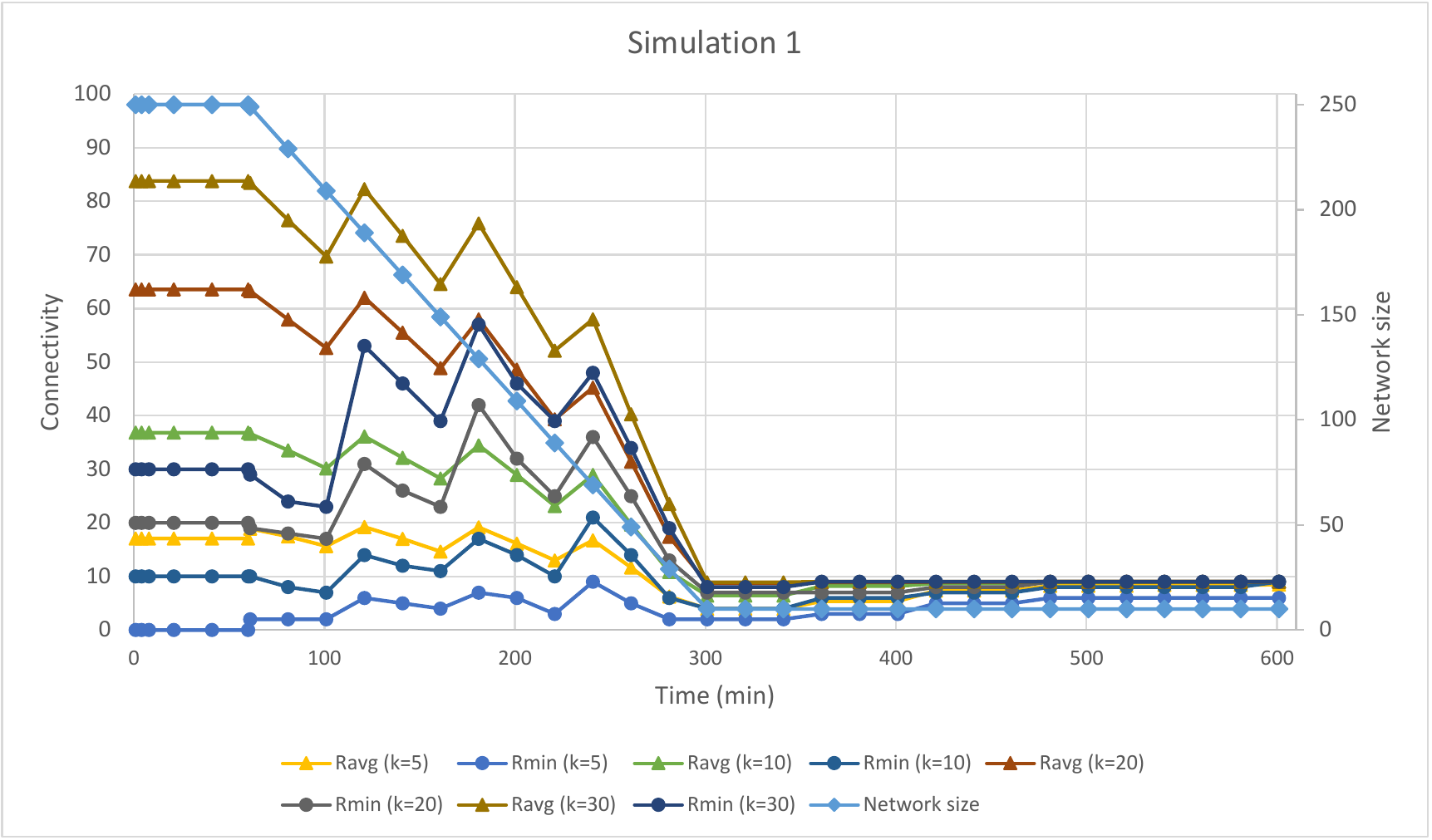}\\
  \caption{Size 250, multiple k, churn 0/1, without traffic}\label{multik:figure:161-164-crop}
\end{figure*}

In this simulation, we used the random scenario for the network setup with a network churn of 0/1 and without network traffic.
The bucket sizes are 5, 10, 20 and 30. We present the results in Figure~\ref{multik:figure:161-164-crop}.
We can observe two effects after the churn starts: 
The minimum connectivity drops between refresh cycles, but increases overall.
We assign this to the circumstance that with decreasing network size, less nodes compete for places in the size limited routing tables.
Also, nodes that left the network are removed from the routing tables and others can take their place.
The churn enables a network restructuring that leads to an increase in the minimum connectivity.
This continues until the network size becomes too small to sustain this behaviour.
Towards the end of the simulation, with 10 nodes left in the network, for each bucket size except for 5, the network becomes fully connected, resulting in a connectivity of 9.

In another simulation with the same parameters but without the churn, Kademlia establishes a minimum connectivity of $k$ for all values above $5$ after bootstrap.
After the first bucket refresh cycle at the 60 minute mark, the connectivity of 5 is also reached with $k=5$.
Afterwards, these levels do not change for the remainder of the simulation time.
We did not include a graph for the simulation without churn here since the results of Simulation 1 in Figure~\ref{multik:figure:161-164-crop} displays this exact behaviour in the first 60 minutes.

\subsubsection*{Simulation 2}

\begin{figure*}
  \centering
  \includegraphics[width=\textwidth]{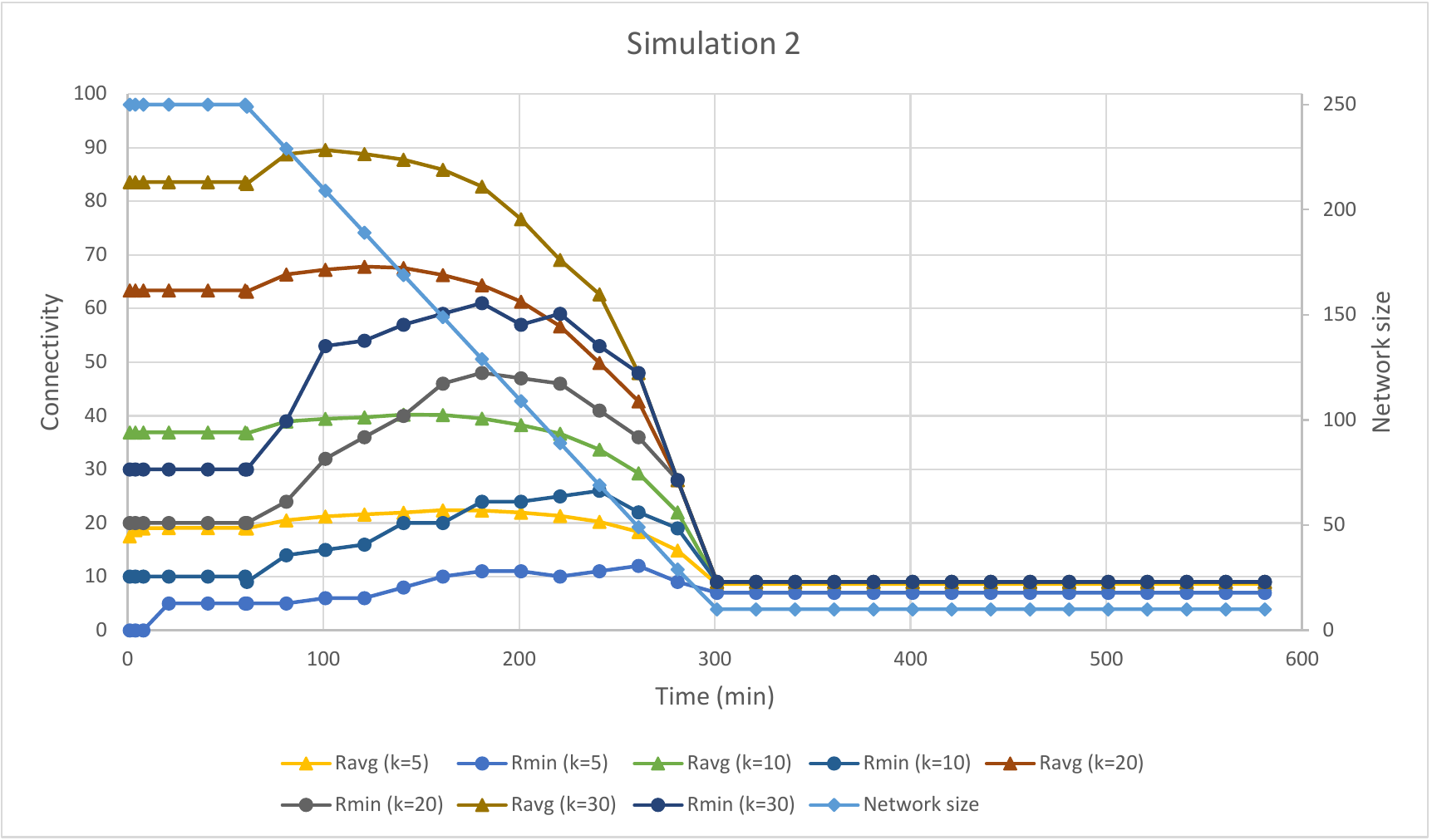}\\
  \caption{Size 250, multiple k, churn 0/1, with traffic}\label{multik:figure:166-169-crop}
\end{figure*}

In this simulation, we used the random scenario for the network setup with a network churn of 0/1 and with network traffic.
The bucket sizes are 5, 10, 20 and 30. We present the results in Figure \ref{multik:figure:166-169-crop}.
In comparison with Simulation 1, the network traffic allows for faster adaptation with respect to removed nodes in the network.
The overall connectivity reaches higher values than in Simulation 1.
Towards the end of the simulation, with 10 nodes left in the network, again for the $k$ values 10 to 30 the network becomes fully connected, while with $k=5$ the connectivity becomes 6.

\subsubsection*{Simulation 3}

\begin{figure}
  \centering
  \includegraphics[width=\columnwidth]{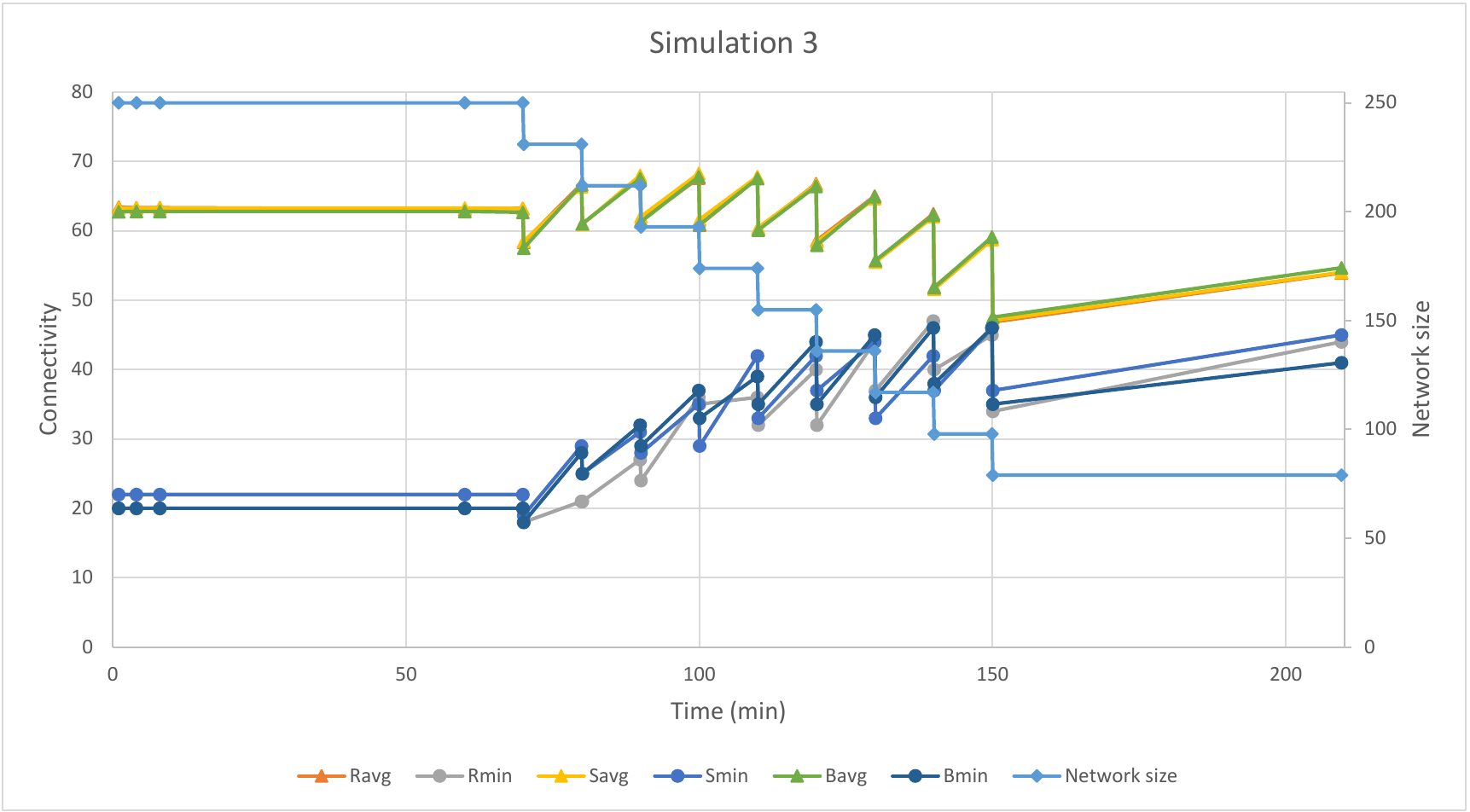}\\
  \caption{Size 250, k=20, churn 0/19, with traffic}\label{multik:figure:144-146-crop}
\end{figure}

In this simulation, we used the random, stable and bootstrap scenarios for network setup with a network churn of 0/19 (burst).
The simulation uses network traffic, and the bucket size is 20. We present the results in Figure~\ref{multik:figure:144-146-crop},
All three scenarios initially reach a connectivity of 20 or 21.
Therefore, a removal of 19 nodes does not disconnect the network.
The network traffic enables sufficient adaptation by the nodes, so that even several burst removals in a row leave the network connected.
With decreasing network size, the connectivity even increases -- an effect we also saw in Simulations 1 and 2.

\subsubsection*{Simulation 4}

\begin{figure}
  \centering
  \includegraphics[width=\columnwidth]{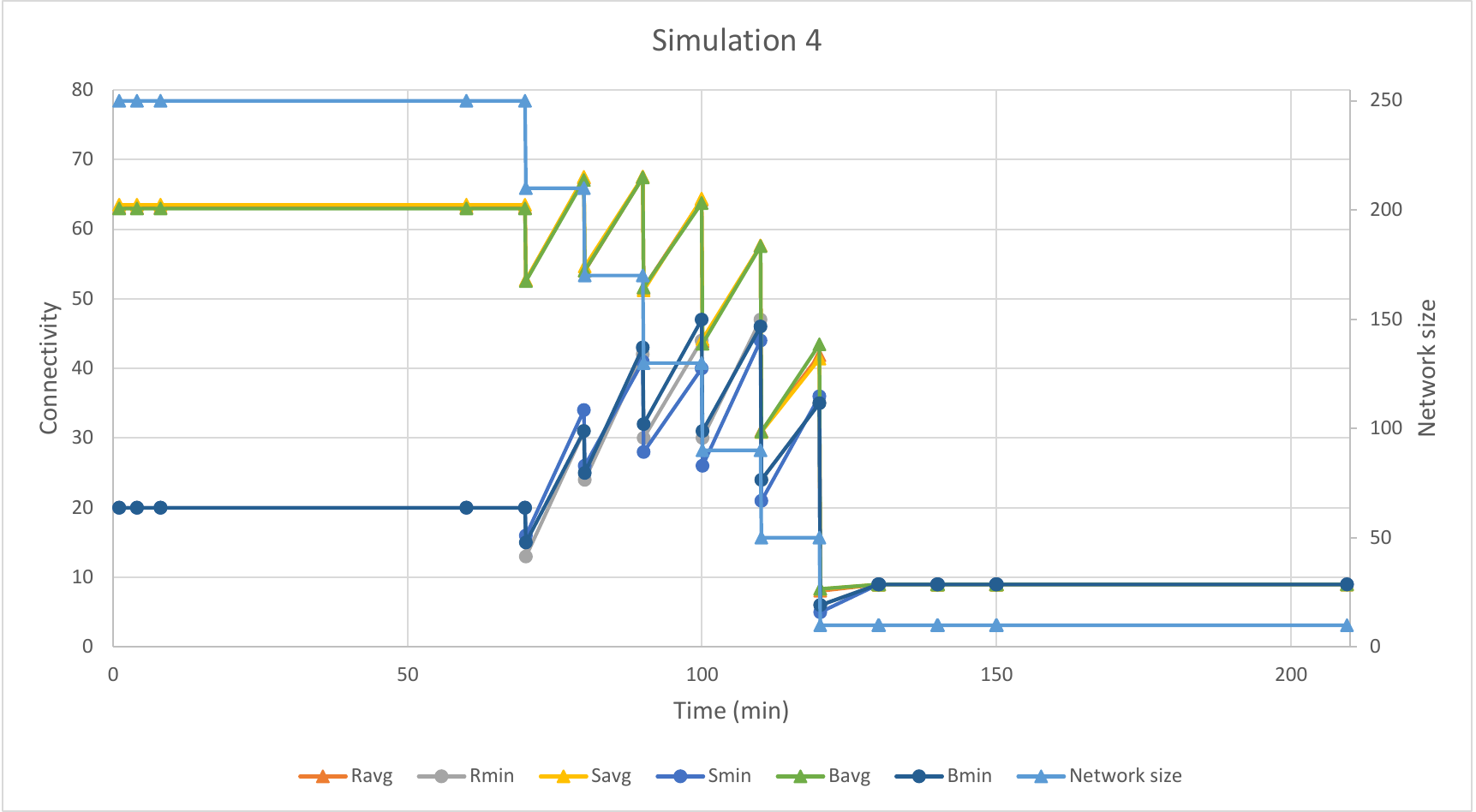}\\
  \caption{Size 250, k=20, churn 0/40, with traffic}\label{multik:figure:147-149-crop}
\end{figure}

In this simulation, we used the random, stable and bootstrap scenario for network setup with a network churn of 0/40 (burst).
The simulation uses network traffic, and the bucket size is 20. We present the results in Figure~\ref{multik:figure:147-149-crop}.
As in Simulation 3, the three scenarios initially reach a connectivity of 20 or 21.
The networks are able to sustain a repeated removal of 40 nodes without becoming disconnected.
This is a significant result: Even for repeated churn of twice the initial connectivity, the network stays connected.

\subsubsection*{Simulation 5}

\begin{figure}
  \centering
  \includegraphics[width=\columnwidth]{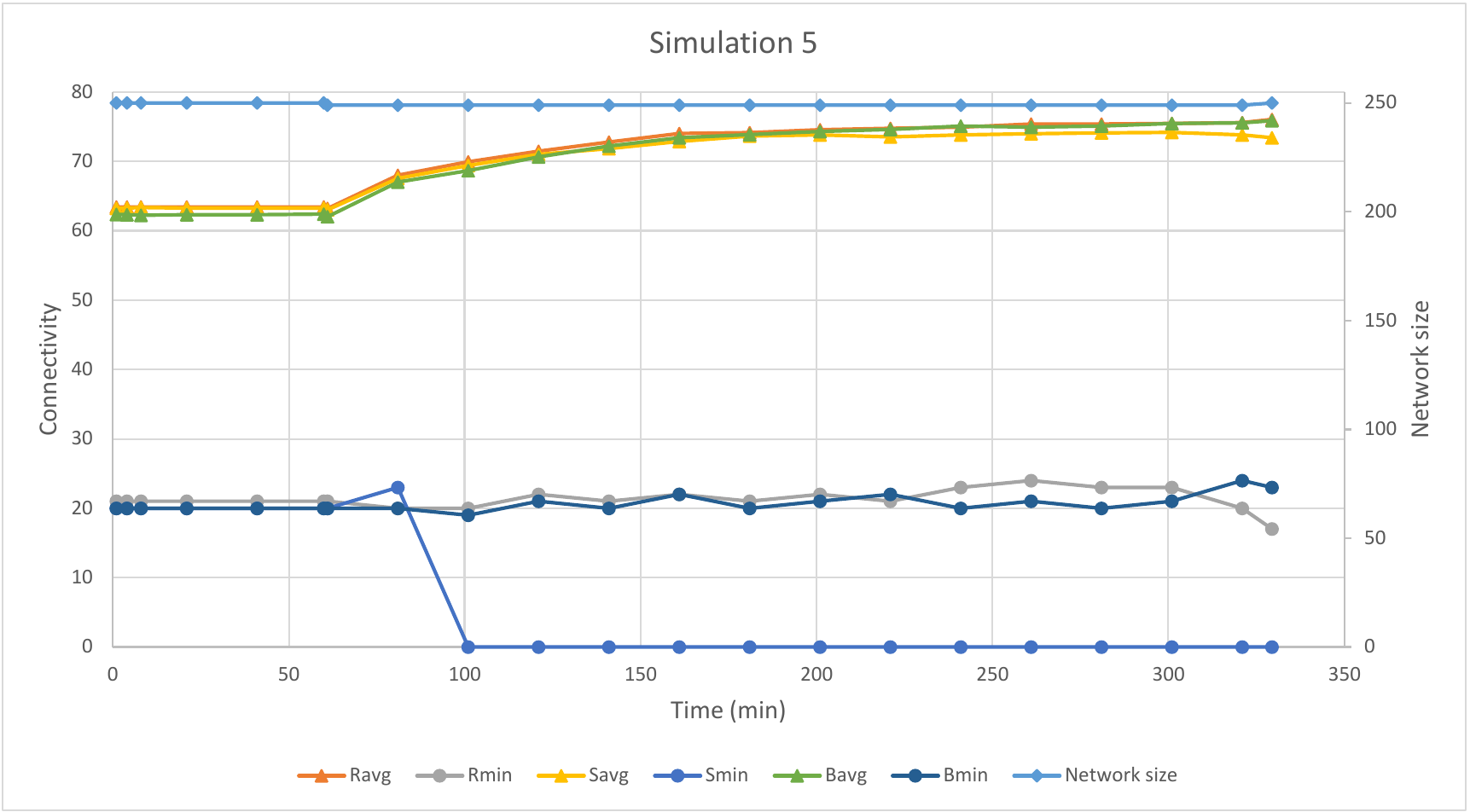}\\
  \caption{Size 250, k=20, churn 1/1, with traffic}\label{multik:figure:141-143-crop}
\end{figure}

In this simulation, we used the random, stable and bootstrap scenario for network setup with a network churn of 1/1.
The simulation uses network traffic, and the bucket size is 20. We present the results in Figure~\ref{multik:figure:141-143-crop}.
In contrast to the previous simulations, now, not only do nodes leave the network, but new nodes join at the same time.
For the stable setup, in one of the five simulation passes a race condition occurred. A node could not join, because its bootstrap node had been removed from the network. In the other four passes, the stable setup was comparable to the random setup in the graph.
Overall the bootstrap setup seems to handle the churn best. After $280$ minutes of churn, it retains the highest connectivity as well as the highest average connectivity.

\subsubsection*{Simulation 6}

\begin{figure}
  \centering
  \includegraphics[width=\columnwidth]{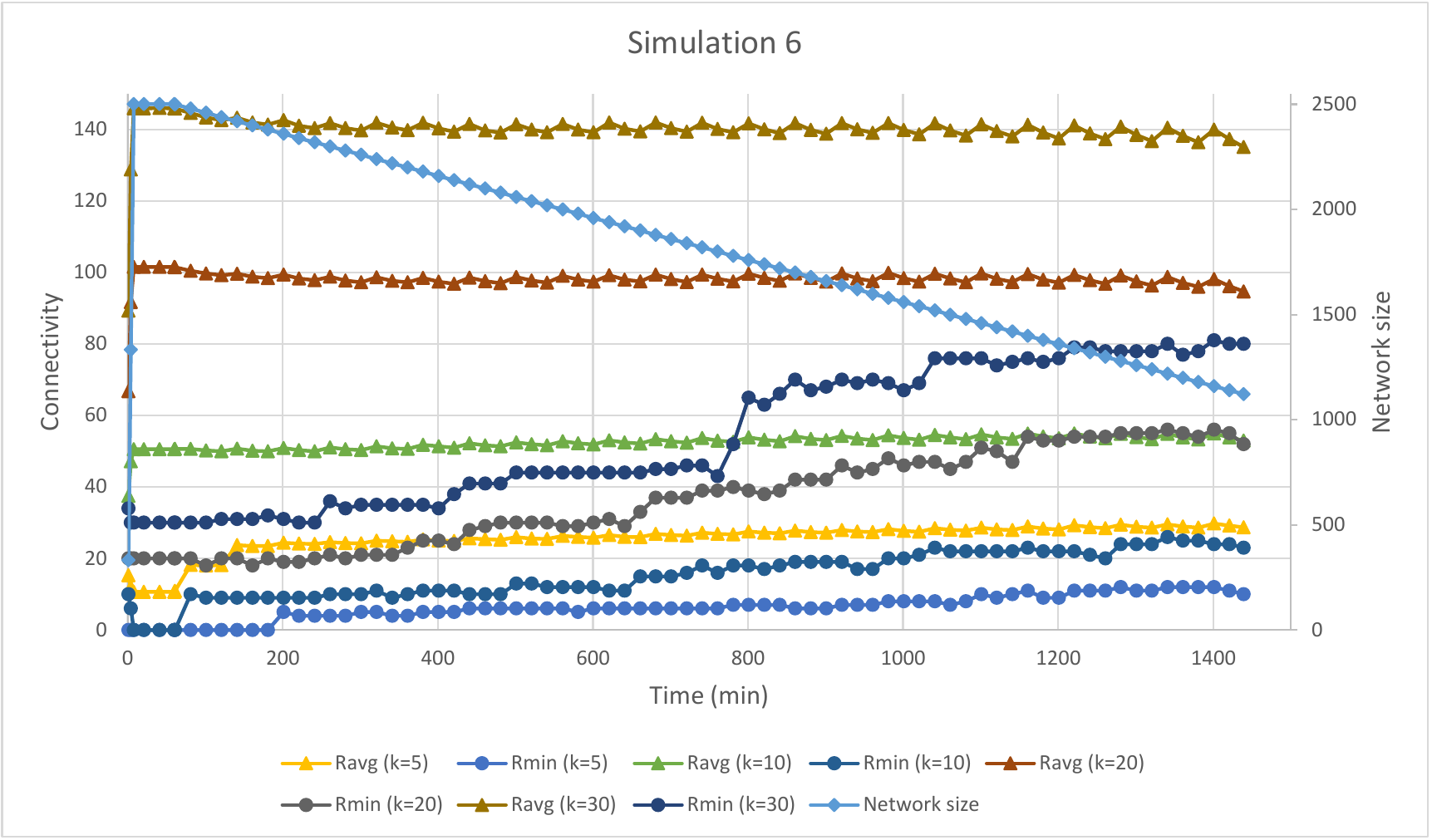}\\
  \caption{Size 2500, multiple k, churn 0/1, without traffic}\label{multik:figure:101-104-crop}
\end{figure}

In this simulation, we used the random scenario for network setup with a network churn of 0/1 and without network traffic.
The bucket sizes are 5, 10, 20 and 30. We present the results in Figure \ref{multik:figure:101-104-crop}.
This simulation is analog to Simulation 1, but with a network size of 2500 nodes.
While in Simulation 1 the network was disconnected for bucket size 5 at the beginning, it is disconnected in this simulation for both bucket sizes $k=5$ and $k=10$.
For the larger network, these bucket sizes seem too small to achieve a connected graph already during the network setup.
Before the churn starts, the connectivity is close or equal to the bucket size.
During the churn, we again observe increasing connectivity in the network with decreasing network size.
For the bucket size 20, the connectivity reaches 50, and, for the bucket size 30, it reaches 80 towards the end of the simulation.

\subsubsection*{Simulation 7}

\begin{figure}
  \centering
  \includegraphics[width=\columnwidth]{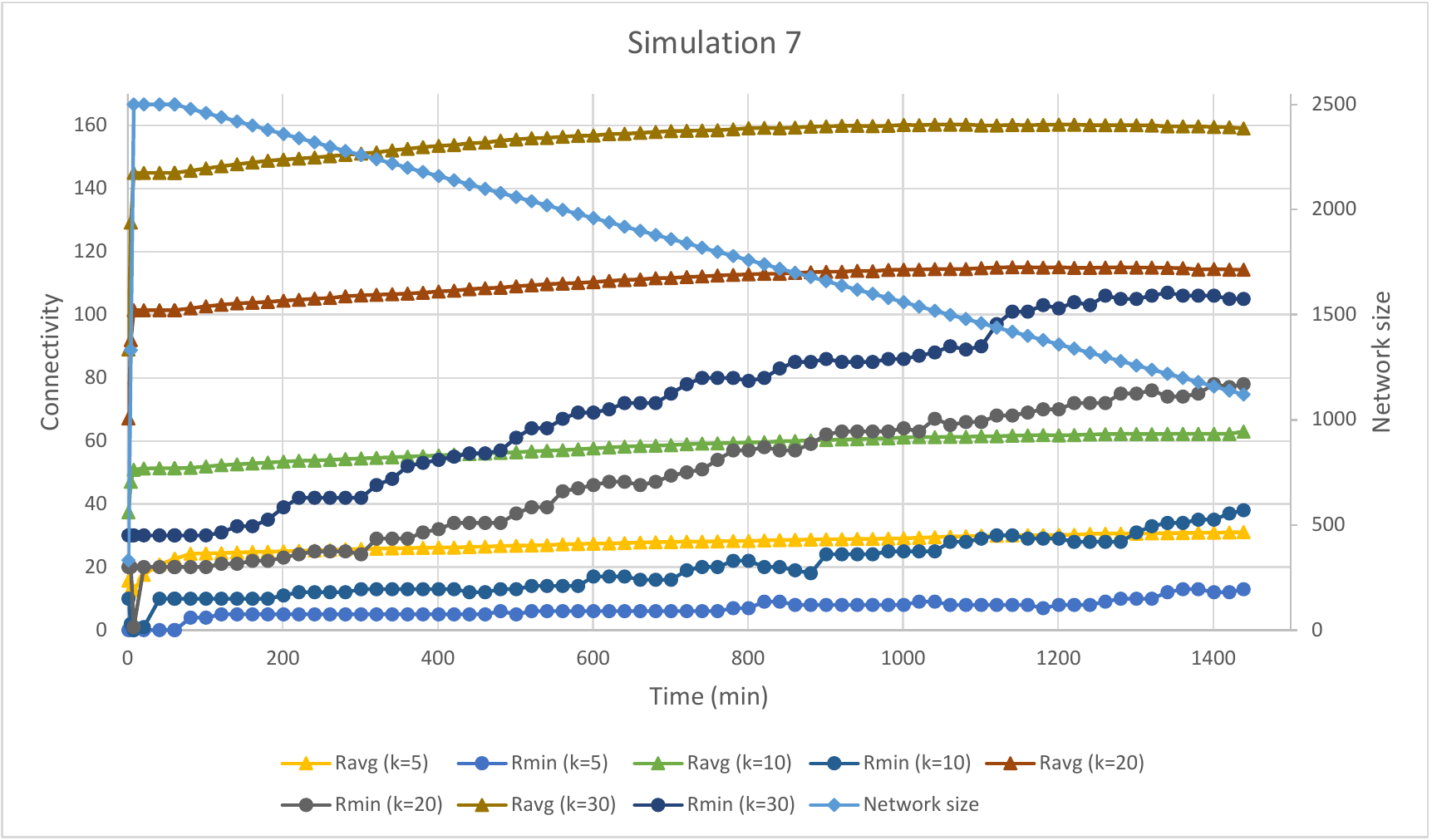}\\
  \caption{Size 2500, multiple k, churn 0/1, with traffic}\label{multik:figure:106-109-crop}
\end{figure}

In this simulation, we used the random scenario for network setup with a network churn of 0/1 and with network traffic.
The bucket sizes are 5, 10, 20 and 30. We present the results in Figure \ref{multik:figure:106-109-crop}.
In comparison to Simulation 6, the traffic enables faster adaptation to the churn and provides higher connectivity.
Towards the end of the simulation, the connectivity for the bucket size 20 reaches 80, and, for the bucket size 30, it reaches 105.

\subsubsection*{Simulation 8}

\begin{figure}
  \centering
  \includegraphics[width=\columnwidth]{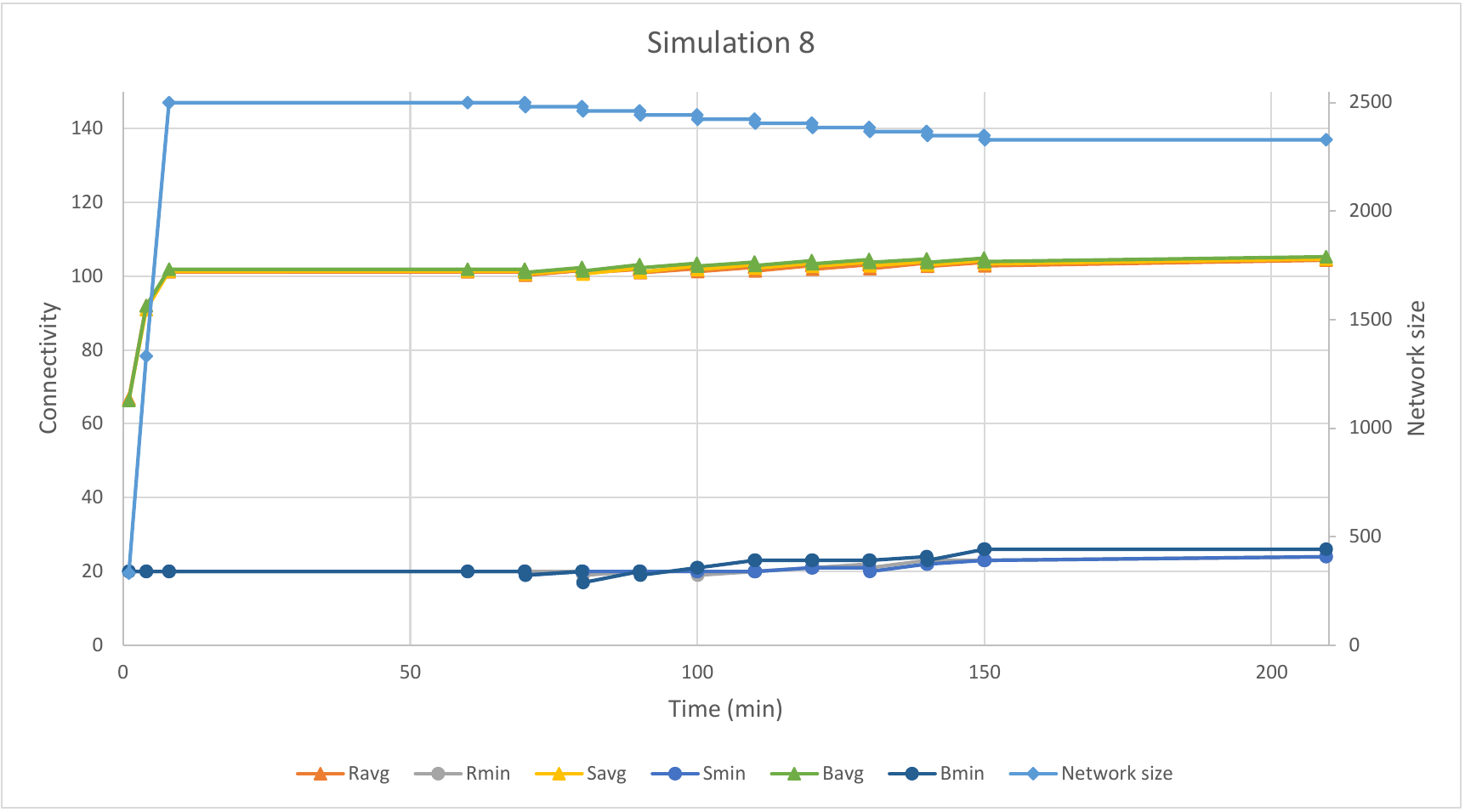}\\
  \caption{Size 2500, k=20, churn 0/19, with traffic}\label{multik:figure:084-086-crop}
\end{figure}

In this simulation, we used the random, stable and bootstrap scenario for network setup with a network churn of 0/19 (burst).
The simulation uses network traffic, and the bucket size is 20. We present the results in Figure~\ref{multik:figure:084-086-crop}.
While in Simulation 3 each churn burst led to a notable reduction of the connectivity, here, the reduction is at most one.
On the other hand, this reduced impact is also notable with the overall increase of the connectivity, which is almost non-existent.

\subsubsection*{Simulation 9}

\begin{figure}
  \centering
  \includegraphics[width=\columnwidth]{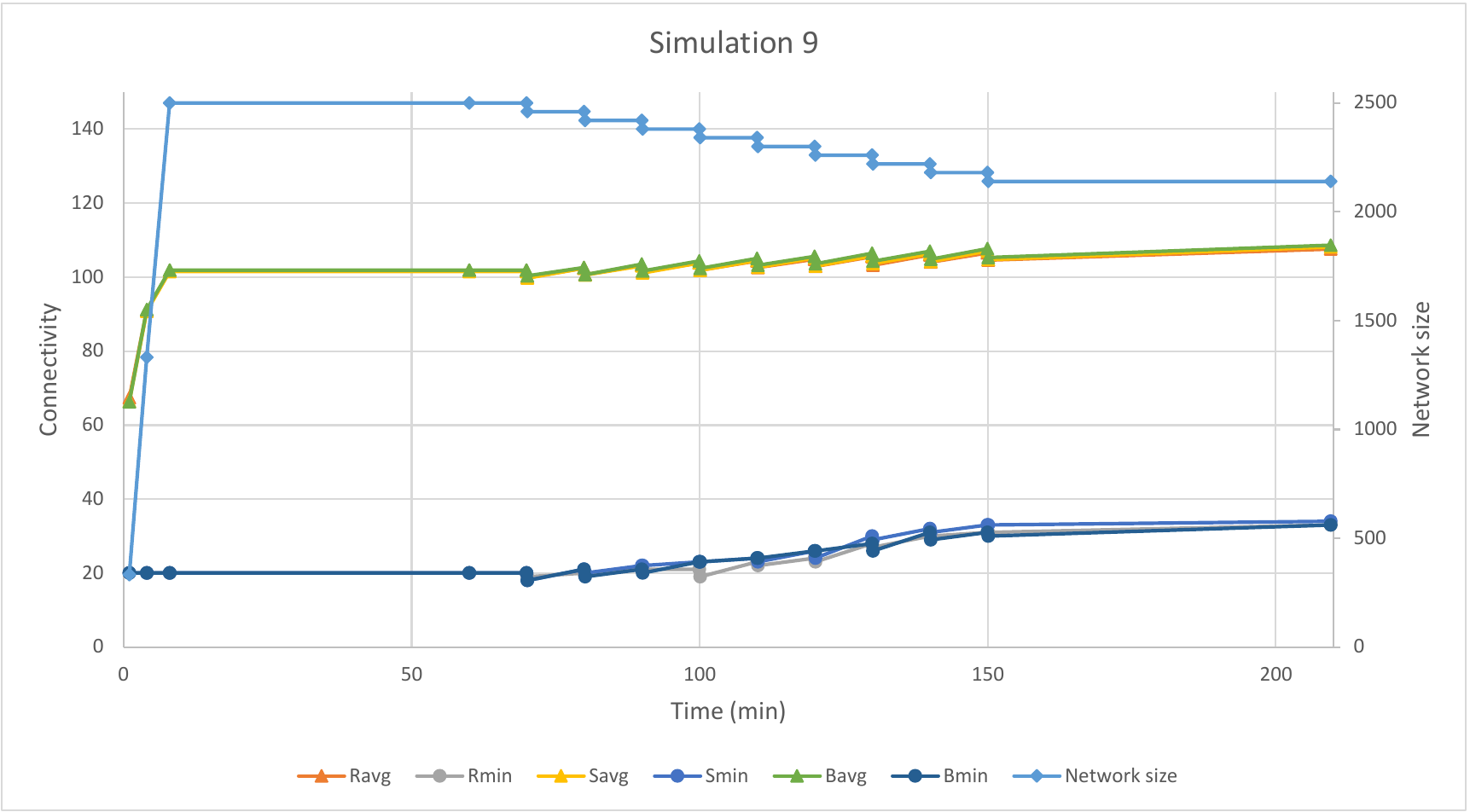}\\
  \caption{Size 2500, k=20, churn 0/40, with traffic}\label{multik:figure:087-089-crop}
\end{figure}

In this simulation, we used the random, stable and bootstrap scenario for network setup with a network churn of 0/40 (burst).
The simulation uses network traffic, and the bucket size is 20. We present the results in Figure~\ref{multik:figure:087-089-crop}.
As with Simulations 3 and 8, the effect of repeated churn bursts of twice the bucket size is significantly less than in Simulation 4.
With each churn burst, the connectivity drops by two.
In total, the connectivity increases more strongly than in Simulation 8, which is probably due to a stronger decrease in the network size.

\subsubsection*{Simulation 10}

\begin{figure}
  \centering
  \includegraphics[width=\columnwidth]{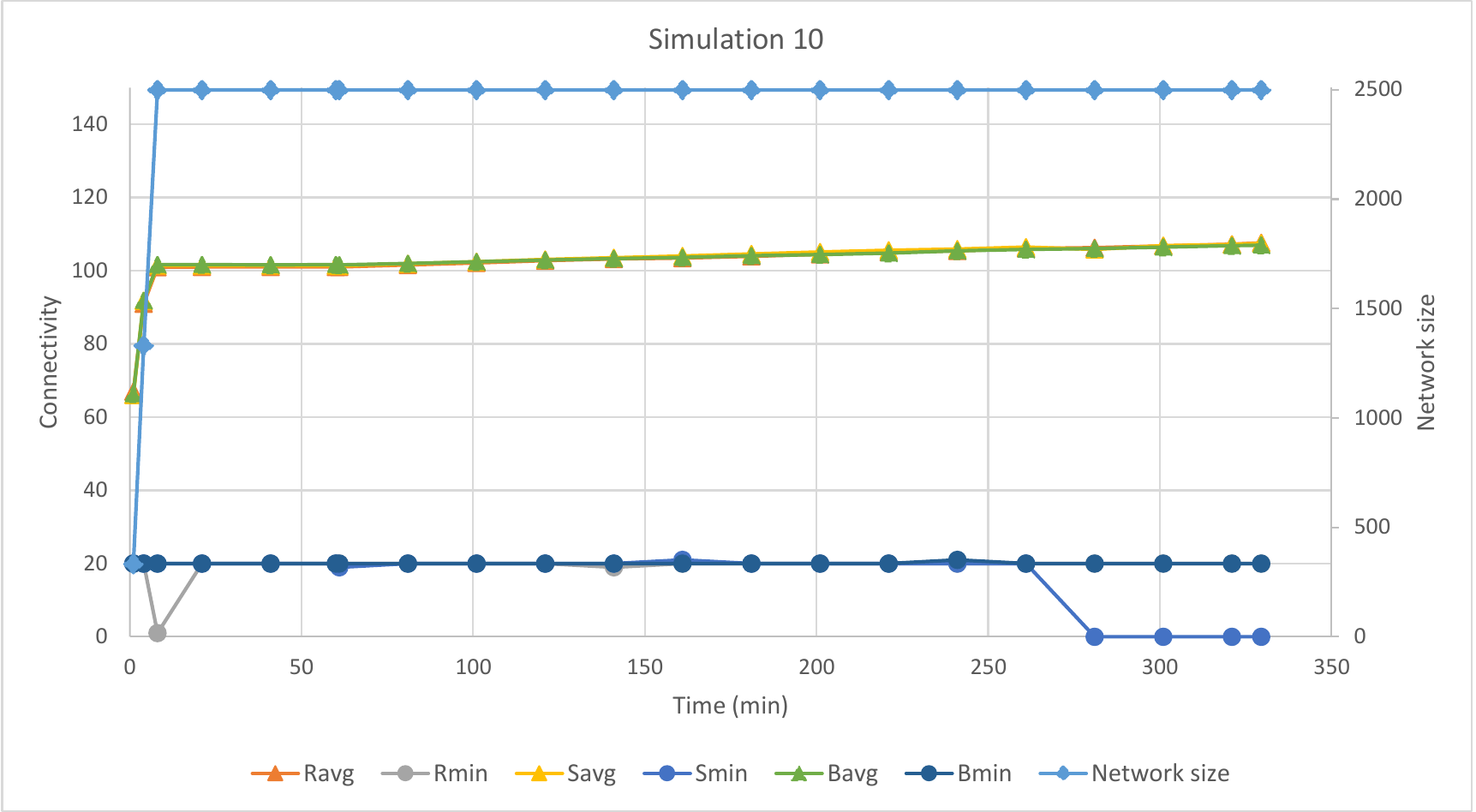}\\
  \caption{Size 2500, k=20, churn 1/1, with traffic}\label{multik:figure:081-083-crop}
\end{figure}

In this simulation, we used the random, stable and bootstrap scenario for network setup with a network churn of 1/1.
The simulation uses network traffic, and the bucket size is 20. We present the results in Figure~\ref{multik:figure:081-083-crop}.
Over its runtime, this simulation shows a connectivity similar to that in Simulation 5.
Though the network size is ten times as large, the connectivity under 1/1 churn remains near the initial value of 20.
At minute 8, just at the end of the initial network setup, the connectivity drops to one for the random scenario.
At the time of the next connectivity snapshot, the network has recovered, and the connectivity is back at 20.
At minute 281, the connectivity of the stable scenario drops to zero.
As with Simulation 5, this is due to a race condition in the join process and happened in one of the five simulation passes.

\subsubsection*{Simulation 11}

\begin{figure}
  \centering
  \includegraphics[width=\columnwidth]{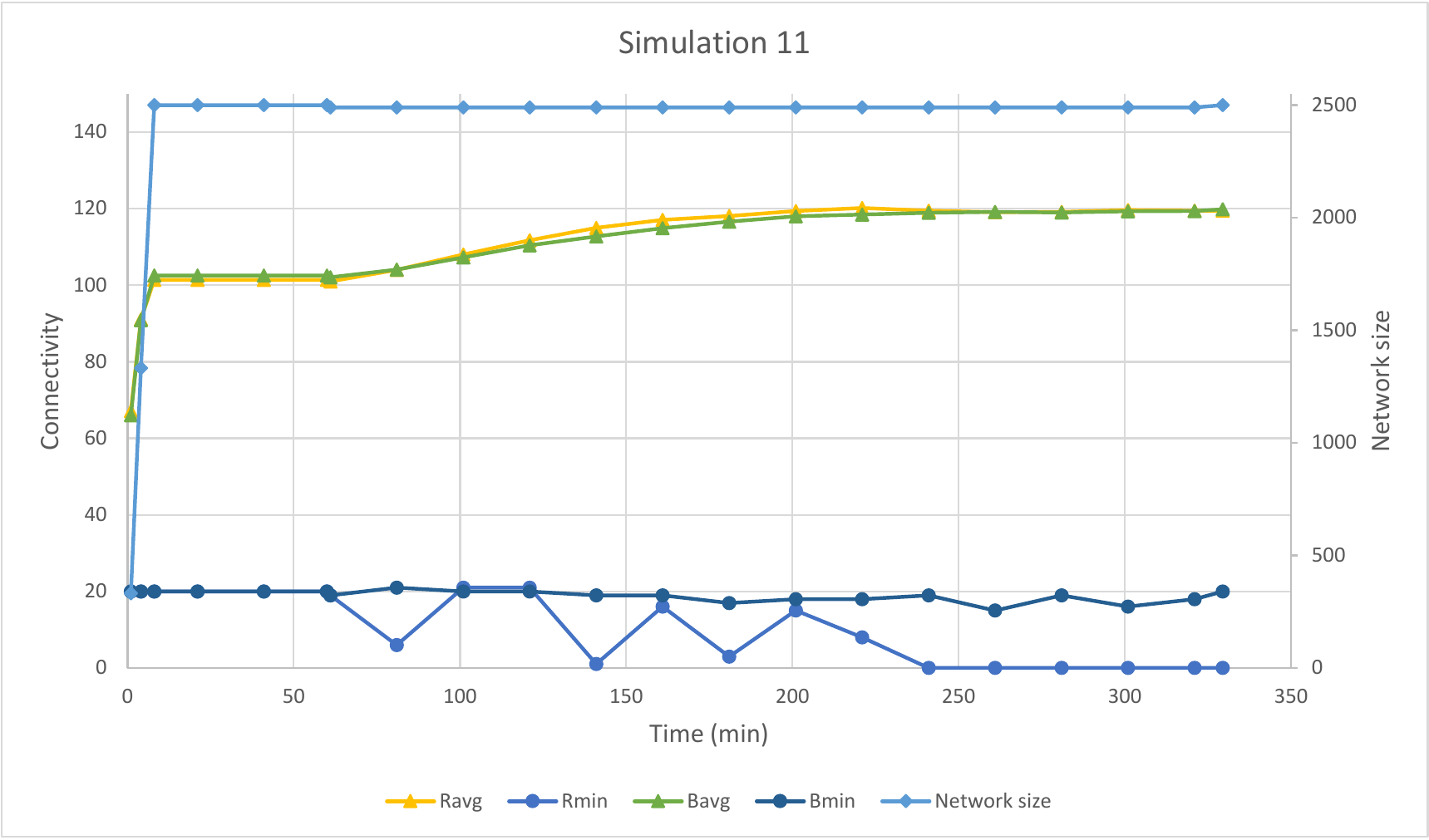}\\
  \caption{Size 2500, k=20, churn 10/10, with traffic}\label{multik:figure:115-116-crop}
\end{figure}

In this simulation, we used the random and the bootstrap scenario for network setup with a network churn of 10/10.
The simulation uses network traffic, and the bucket size is 20. We present the results in Figure \ref{multik:figure:115-116-crop}.
The previous Simulations 5 and 10 have shown connectivity problems for either the random or the stable scenario.
To better evaluate the effect of this kind of churn, we performed an additional simulation with 10/10 churn.
As representative of the setups with random selection of bootstrap nodes we chose the random setup and compared it to the bootstrap setup.

The random scenario shows drops in connectivity several times during the simulation:
it drops to 6 at minute 81, to 1 at minute 141, to 3 at minute 181, and to 8 at minute 221.
Shortly after minute 221, the network becomes disconnected and does not recover for the remainder of the simulation.
The bootstrap scenario oscillates around the initial connectivity of 20, but does neither experience significant drops in connectivity nor becomes disconnected.

\section{Conclusion \& Future Work}
\label{conn:sec:conclusion}

In this paper, we analyzed the communication resilience of a highly distributed self-organizing CPS.
Such a system exhibits coordination schemata and communication requirements similar to structured overlay networks.
To achieve reliable self-adaption, we require redundant communication channels for resilient inter-node communication.
Specifically, we used Kademlia as the communication overlay for the CPS and analyzed its network connectivity by simulations.

Our main result from the simulations is that the network connectivity $\kappa$ of Kademlia strongly correlates with its bucket size $k$.
In most simulations, the network connectivity was equal or higher than $k$.
There are also cases, especially with high churn, when the network connectivity drops significantly under $k$.
However, to achieve a certain resilience level $r$ in a CPS, we require a network connectivity $\kappa > r$. With our results, we determined that the bucket size needs to be set to a value higher than $r$, i.e., $k>r$. Nevertheless, the resilience level cannot be guaranteed.

In the future, we plan to extend Kademlia to guarantee the network connectivity in all cases. Even more, we plan to introduce a parameter into Kademlia for controlling its connectivity independently of the bucket size.

\section*{Acknowledgment}
The authors would like to thank the German Research Foundation (DFG) for support within the DFG project CYPHOC (WA 2828/1-1).

\bibliographystyle{abbrv}

\bibliography{connection}

\end{document}